# Testing separability and independence of perceptual dimensions with general recognition theory: A tutorial and new R package (grtools)

Fabian A. Soto[1], Emily Zheng[2], Johnny Fonseca[3], & F. Gregory Ashby[4]

[1]Department of Psychology, Florida International University
[2]Department of Statistics and Applied Probability, University of California, Santa Barbara
[3]Department of Mathematics and Statistics, Florida International University
[4]Department of Psychological & Brain Sciences, University of California, Santa Barbara

## Abstract

Determining whether perceptual properties are processed independently is an important goal in perceptual science, and tools to test independence should be widely available to experimental researchers. The best analytical tools to test for perceptual independence are provided by General Recognition Theory (GRT), a multidimensional extension of signal detection theory. Unfortunately, there is currently a lack of software implementing GRT analyses that is ready-to-use by experimental psychologists and neuroscientists with little training in computational modeling. This paper presents *grtools*, an R package developed with the explicit aim of providing experimentalists with the ability to perform full GRT analyses using only a couple of command lines. We describe the software and provide a practical tutorial on how to perform each of the analyses available in *grtools*. We also provide advice to researchers on best practices for experimental design and interpretation of results.

In perceptual science, an important amount of effort has been dedicated to understanding what aspects of stimuli are represented independently (e.g., Bruce and Young, 1986; Haxby et al., 2000; Kanwisher, 2000; Op de Beeck et al., 2008; Stankiewicz, 2002; Ungerleider and Haxby, 1994; Vogels et al., 2001). Independent processing of two stimulus properties is interesting, because it implies that those properties are given priority by the perceptual system, which allocates a specific set of resources to the processing of each. On the other hand, some properties are said to be processed "holistically" or "configurally," which is equivalent to saying that they cannot be processed independently (e.g., Mestry et al., 2012; Richler et al., 2008; Thomas, 2001). Such holistic processing is also important to understand perception, as it might be an indicator of perceptual specialization or expertise (see Farah et al., 1998; Maurer et al., 2002; Richler et al., 2012). In sum, determining whether perceptual properties are processed independently is an important goal in perception science, and tools to test independence should be widely available to experimental researchers.

Currently, the best analytical tools to test for perceptual independence are provided by General Recognition Theory (GRT; Ashby and Townsend, 1986; for a tutorial review, see Ashby and Soto, 2015). GRT is an extension of signal detection theory to cases in which stimuli vary along two or more dimensions. It offers a framework in which different types of dimensional interactions can be defined formally and studied, while inheriting

This work was supported in part by grant no. W911NF-07-1-0072 from the U.S. Army Research Office through the Institute for Collaborative Biotechnologies, and by NIH grant 2R01MH063760.

from signal detection theory the ability to dissociate perceptual from decisional sources for such interactions. Still, the implementation of GRT analyses requires an important amount of technical knowledge, so this task has been almost exclusively performed by computational modelers. The lack of software implementing GRT analyses that is ready-to-use by untrained researchers has probably reduced the application of GRT among experimental psychologists and neuroscientists.

This paper presents *grtools*, an R package developed with the explicit aim of providing experimentalists with the ability to perform full GRT analyses using only a couple of command lines. We describe the software and provide a practical tutorial on how to perform each of the analyses available in *grtools*. The goal is to give a step-by-step guide for researchers interested in applying GRT to their own research problems. Readers interested in the theory behind these analyses should consult the recent review by Ashby and Soto (2015) and the papers referenced therein.

## 1. General Recognition Theory

GRT is a multivariate extension of signal detection theory to cases in which stimuli vary on more than one dimension (Ashby and Townsend, 1986; for a review, see Ashby and Soto, 2015). As in signal detection theory, GRT assumes that different presentations of the same stimulus produce slightly different perceptual representations. However, because stimuli vary in multiple dimensions, perceptual representations vary along multiple dimensions at the same time. For example, imagine that you are interested in the dimensions of gender (male vs. female) and emotional expression (neutral vs. sad) in faces. A single presentation of a happy female (Figure 1a) would produce a perceptual effect, a single point in the two-dimensional space of perceived gender and emotional expression. Figure 1b shows an example of such a two-dimensional perceptual space; each point in the figure represents a percept evoked by a single stimulus presentation, and the tick marks on the axes represent the corresponding percept values on both perceptual dimensions. The green dotted lines show this correspondence between points and tick marks more clearly for a single perceptual effect.

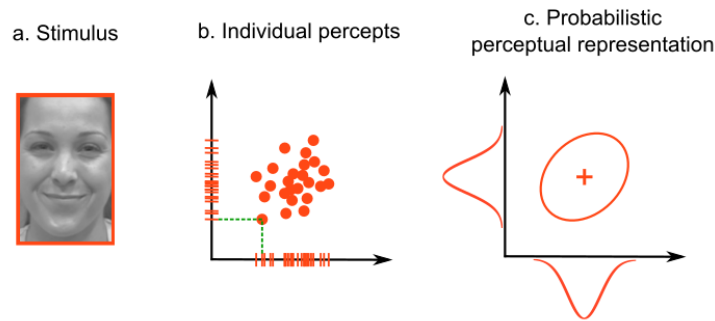

**Figure 1.** Stimulus representation in general recognition theory. For (a) faces varying in gender (male or female) and emotional expression (sad or happy), the representation is two-dimensional. Each stimulus presentation produces (b) a point in the two-dimensional space. The whole distribution of such perceptual effects can be described through (c) a two-dimensional probability distribution.

Thus, the representation of a happy female face in Figure 1 is probabilistic across trials and can be summarized through a probability distribution. Assuming that this distribution is a two-dimensional Gaussian, we can represent it graphically by the ellipse in Figure 1c, which is the shape of the cloud of points that are produced by the left presented face. The plus sign inside the ellipse represents the mean of the distribution. The bell-shaped unidimensional Gaussians plotted on the axes of Figure 1c represent the distribution of tick marks along a single dimension. These are called *marginal distributions*.

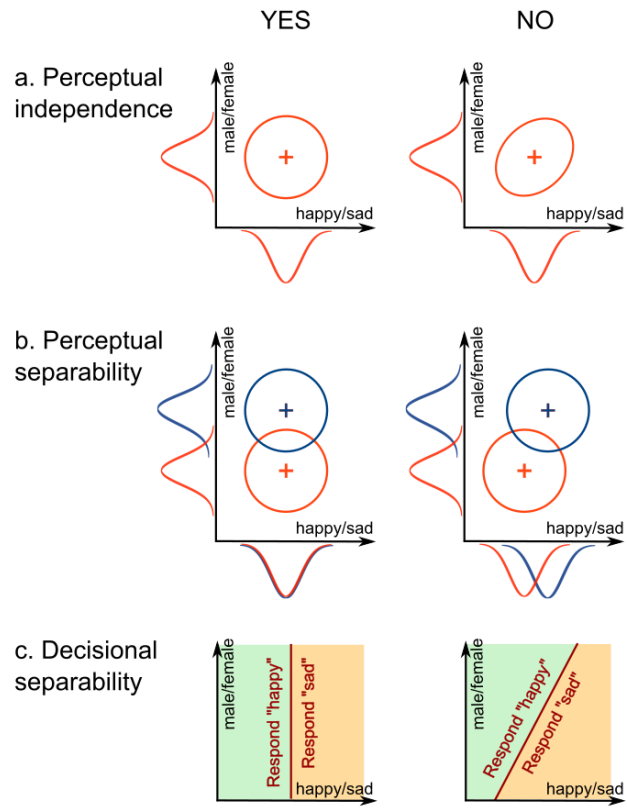

**Figure 2.** Schematic representation of different forms of interaction between dimensions defined in general recognition theory: (a) perceptual independence, (b) perceptual separability, and (c) decisional separability.

Figure 2a shows two distributions that are equivalent except for the correlation between the values of percepts on the two dimensions. The distribution to the left shows no correlation; this is a case of *perceptual independence*, where the perceived value of expression does not depend on the perceived value of gender. The distribution to the right shows a positive correlation, which is an example of a *failure of perceptual independence*. In this case, whenever the face is presented, the more it is perceived as female-looking, the more likely it is that it will also be perceived as happier. It is important to remember that perceptual independence is defined for a single stimulus.

Another important concept in GRT is related to how two or more stimuli are perceived. Each stimulus with a unique combination of gender and emotional expression can be represented by its own probability distribution. For example, Figure 2b shows perceptual distributions for two happy faces that differ in their gender. Note how the distributions in the left panel are aligned on the x-axis, representing expression, and how their marginal distributions overlap. This is a case of *perceptual separability* of emotional expression from gender; the perception of happiness is not affected by the face's gender. The right panel of Figure 2b shows an example of the opposite case, a *failure of perceptual separability*. In this case, the two distributions do not align and their marginal distributions do not overlap. The female face is perceived as more "happy" than the male face.

In a typical behavioral experiment, participants would be asked to identify something about the presented stimulus, like a face's expression, gender, or both. According to GRT, participants achieve this by dividing the perceptual space into response regions. For example, a participant asked to report gender might divide the perceptual space as shown in Figure 2c, using a linear bound. When a percept lands on the left area, the participant reports "happy," and when the percept lands on the right area, the participant reports "sad." Note how the decision bound in the left panel of Figure 2c, which is orthogonal to the expression dimension, divides the space in the same way across all values of gender. This is a case of *decisional separability* of emotional expression from gender; the decisions about expression are not affected by the face's gender. The right panel of Figure 2c shows an example of the opposite case, a *failure of decisional separability*. In this case, the bound is tilted instead of orthogonal, and the area given to the "happy" response is much smaller in the "male" end of the identity dimension than in the "female" end. That is, the observer is biased to answer "happy" more often for the female faces than for male faces.

## 2. Installing *grtools*

The grtools package uses the computer programming language R (R Core Team, 2016). R is an open-source statistical software used by many researchers across a variety of scientific fields. R can be downloaded for free at http://cran.rstudio.com.

We highly recommend using RStudio to work with R. RStudio is an integrated development environment (IDE) for R, which provides a more user-friendly experience that includes a package manager, syntax-highlighting editor, tools for plotting, history, debugging, and workspace management. Rstudio is also free and open-source, and it can be downloaded at https://www.rstudio.com/products/rstudio/download/.

The grtools package will require that you have a C++ compiler installed on your computer. For Windows users, the C++ compiler Rtools (http://cran.r-project.org/bin/windows/Rtools/) can be used. For Mac users, a C++ compiler can be installed through Xcode (found in the Apple Application Store). More instructions on how to install these compilers can be found in the grtools webpage: https://github.com/fsotoc/grtools.

To install grtools, start a session in R (or RStudio) and in the console type the following:

```
install.packages("devtools")
devtools::install_github("fsotoc/grtools",
dependencies="Imports")
```

Now grtools is installed and available for use. To have access to grtools functions and analyses, you have to load the package into your current R session. Type the following in the console to load the grtools package:

```
library(grtools)
```

To access documentation that provides explanations and examples for the analyses in grtools, type the following in the console:

```
?grtools
```

## 3. The 2x2 identification task

One of the most widely used tasks to study the independence of stimulus dimensions using GRT is the 2x2 identification task. On each trial of an identification task, a stimulus is presented and it must be identified by pressing a specific response button. Each stimulus must have a value on at least two dimensions or features (that we want to test for independence), A and B. If there are only two values per dimension, we obtain the 2x2 design with stimuli $A_1B_1$, $A_1B_2$, $A_2B_1$ and $A_2B_2$.

For example, consider a 2x2 face identification experiment where the two varying dimensions are face emotional expression (dimension A) and gender (dimension B). Assume that the levels for the emotion dimension are happy ($A_1$) and sad ($A_2$), while the levels for the gender dimension are male ($B_1$) and female ($B_2$). Thus, a 2x2 identification task would create 4 face stimuli: happy-male ($A_1B_1$), sad-male ($A_1B_2$), happy-female($A_2B_1$), and sad female ($A_2B_2$). On a given trial, a participant is shown one of these faces, and must identify the face accordingly. Application of GRT to this task requires that participants make identification errors. Because some stimuli (e.g., faces) are very easy to identify, obtaining identification errors might require manipulating the stimuli to increase task difficulty. Manipulations that increase errors include decreasing image contrast, decreasing presentation times, and increasing stimulus similarity through morphing and other image manipulation techniques.

Due to its simplicity, the 2x2 task is by far the most widely used design in the field. It requires presenting only four stimuli and measuring only four responses. Other designs, such as a 3x3 task (two dimensions, each with three levels), are also possible. However, these tasks are less practical because the considerable learning required by participants means that the experiment usually will require multiple experimental sessions (e.g., five days in Experiment 1 of Ashby and Lee, 1991). The working memory load of a 3x3 task is also taxing, as participants are required to remember 9 unique stimuli and their unique responses. Because of these and other reasons (see Ashby and Soto, 2015; Soto et al., 2015), functions in the *grtools* package were developed specifically to deal with the 2x2 identification task.

Some data gathered through the 2x2 identification task must be excluded to allow accurate analyses using *grtools*. First, data from participants in the first few blocks of trials must usually be excluded. These data represent the participants' learning periods for the task. GRT models only decisional and perceptual processes during steady-state performance; it is unable to model learning processes. A good approach is to set a learning criterion (e.g., a specific number of blocks finished, or a percentage of correct responses within the last *n* trials) and use data only after the criterion has been reached by a participant. Second, data from participants that perform at chance levels in the task should also be excluded. This is usually an indicator that the participant did not understand or learn the task well. Lastly, data from participants that perform near-perfectly should also be excluded. The GRT analyses included in *grtools* extract information about dimensional independence from the pattern of errors shown by participants. If there are only a few errors, *grtools* analyses will be misleading and inaccurate.

The data from an identification experiment is summarized in a confusion matrix, which contains a row for each stimulus and a column for each response. In the 2x2 design, there are 4 stimuli and 4 responses, resulting in a 4 x 4 confusion matrix with a total of 16 data elements for each test participant. An example from our hypothetical experiment dealing with face gender and emotion is shown in Table 1. The entry in each cell of this matrix represents the number of trials in which a particular stimulus (row) was presented and a particular response (column) was given. Thus, entries on the main diagonal represent the frequency of each correct response and off-diagonal entries describe the various errors (or confusions). For example, the number in the top-left in Table 1 represents the number of trials on which stimulus A1B1 was presented and correctly identified (140 trials). The next cell to the right shows the number of trials on which stimulus A1B1 was presented, but the participant incorrectly reported seeing stimulus A2B1 (36 trials). Each row of values must sum up to the total number of trials on which the corresponding stimulus was presented.

**Table 1.** Data from a simulated identification experiment with four face stimuli, created by factorially combining two levels of emotion (happy and sad) and two levels of gender (male and female).

| Stimulus | Response | | | |
|---|---|---|---|---|
| | **Happy/Male** | **Sad/Male** | **Happy/Female** | **Sad/Female** |
| Happy/Male | 140 | 36 | 34 | 30 |
| Sad/Male | 89 | 91 | 4 | 66 |
| Happy/Female | 85 | 5 | 90 | 70 |
| Sad/Female | 20 | 59 | 8 | 163 |

When entering the data to R, be careful to order rows and columns in the same way as shown in Table 1; that is, beginning with the first row/column and ending with the last row/column, the correct order is: A1B1, A2B1, A1B2, and A2B2.

There are at least two ways in which the data from a confusion matrix can be entered into R for analysis with *grtools*. The first option is to directly enter the data as an R object in the matrix class. This can be done through the R function `matrix`, as in the following example:

```
cmat <- matrix(c(140, 36, 34, 40, 89, 91, 4, 66, 85, 5, 90,
70, 20, 59, 8, 163), nrow = 4, ncol = 4, byrow = TRUE)
```

Note that we have entered the data as a single vector inside the function `c()`, starting with the data from the first row, then the second row, and so on. The function `matrix` is then used to shape this vector into an actual confusion matrix.

The second option is to input the confusion matrix into a spreadsheet application, such as Microsoft Excel. As before, the data must be ordered into a 4x4 matrix, as shown in Table 1. After entering the data, go to the "File" menu and choose "Save as…". In the drop-down menu titled "Format" choose "Comma Separated Values (.csv)." Name your file, choose the folder where you want to store it, and click "Save." In our example, we will name the file "data.cvs" and store it in the directory "home/Documents".

To import the data to R, make sure first that your working directory is the folder that contains your data file. This is easy to do in RStudio, where you can simply navigate through your folders using the "Files" panel at the bottom-right of the screen. Once you get to your destination folder, click in the "More" menu of the "Files" panel, and choose "Set as Working Directory" (see Figure 3).

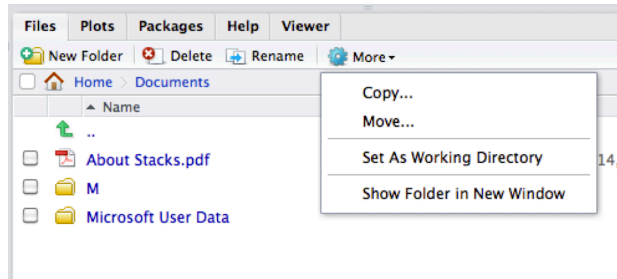

**Figure 3.** Setting the working directory in RStudio.

Read the file using the function `read.table`, as in the following example:

```
cmat <- read.table(file="data.csv", sep=",")
```

Here, the `sep=","` argument specifies the separator used in the data file, which is a comma. Alternatively, if you know the full path of your file, you can include it in the `file` argument (e.g., `file="/home/Documents/data.csv"`) and skip the step of changing the working directory.

The data table is now available as a data frame named `cmat`. The final step is to convert the data into a matrix object, using the `as.matrix` function. The following command reassigns `cmat` as a matrix:

```
cmat <- as.matrix(cmat)
```

### 3.1 Summary statistics analysis

Through a summary statistics analysis, one can draw inferences about perceptual independence, perceptual separability, and decisional separability by using summary statistics that are easily computed from a confusion matrix. An introductory tutorial about the exact statistics computed in this analysis can be found elsewhere (Ashby and Soto, 2015). Here, we focus simply on how to obtain the results from such analysis using *grtools* and how they should be interpreted. For a rigorous treatment of the theory behind these analyses, see the following references: (Ashby and Townsend, 1986; Kadlec and Townsend, 1992a, 1992b).

There are two types of summary statistics analyses that can be performed for the 2x2 identification design: macro- and micro-analyses (Kadlec and Townsend, 1992a, 1992b). Both depend on the computation of only three statistics—proportion of correct responses, sensitivity, and bias— for different combinations of stimuli and responses.

A complete set of macro-analyses can be performed in *grtools* with only three lines of code executed in the R console. The first line performs the actual analysis, using the data that we previously stored in the matrix named `cmat`:

```
macro_results <- sumstats_macro(cmat)
```

This stores the results in an object named `macro_results`, of class `sumstats_macro`. Our second line of code allows us to see a summary of the results:

```
summary(macro_results)
```

This should produce an output table like the one reproduced in Figure 4a. The interpretation of this table is straightforward. Each row represents a particular dimension, and the columns titled "MRI" (Marginal Response Invariance), "Marginal $d'$" and "Marginal $c$" include information about whether or not each of these conditions holds according to the statistical tests performed (see Ashby and Soto, 2015). Only two values will be displayed in these columns: "YES" to indicate that the condition holds, and "NO" to indicate that the condition does not hold (i.e., a significant failure was detected in the relevant test).

**Figure 4.** (a) Summary and (b) full results of a macro-analysis of data from a 2x2 identification design with grtools.

Information about the analyses' conclusions is stored in the final two columns, named "PS" for perceptual separability, and "DS" for decisional separability. The results in these columns are formatted similarly to the results from Kadlec's classic msda2 software (Kadlec, 1995; Table 1):

- yes means that PS/DS may hold (weak evidence).
- NO means that PS/DS does not hold.
- ?yes means that PS/DS is unknown but possibly yes
- ?no means that PS/DS is unknown but possibly no
- ? means that PS/DS is unknown

In our specific example (Figure 4a), we conclude that dimension A is not perceptually separable from dimension B (NO), but dimension B may be perceptually separable from dimension A (yes). We also conclude that dimension B may be decisionally separable from dimension A (yes), but there is not enough information to reach a conclusion about whether dimension A is decisionally separable from dimension B (?).

If we want to obtain more specific information about the tests performed in the macro-analysis, we simply type the name of the object in which we stored our results:

```
macro_results
```

This produces the output in Figure 4b. Results from each of the tests summarized in Figure 4a are now displayed in full detail, including the specific sub-tests, computed statistics, *p*-values and conclusions.

Micro-analyses are another method of summary statistic analyses available in *grtools* (see Ashby and Soto, 2015). These analyses use a different set of summary statistics to test assumptions about both perceptual independence and separability. The sequence of commands required to perform a full set of micro-analyses is similar to the macro-analyses described above. The first line performs the micro-analyses using the data stored in the matrix `cmat`:

```
micro_results <- sumstats_micro(cmat)
```

This stores the results in an object named `micro_results`. To see a summary of the results, you can use this `micro_results` object as input to the `summary` function:

```
summary(micro_results)
```

This produces an output table like the one reproduced in Figure 5a. Each row represents a unique stimulus, and the columns titled "Sampling Independence", "Equal Cond d' " (Equal Conditional d'), "Equal Cond c" (Equal Conditional c), include information about whether or not each of these conditions holds according to the statistical tests performed (Ashby and Soto, 2015). Conclusions regarding perceptual independence and decisional separability are stored in the final two columns, labeled PI and DS respectively. The format of the results is the same as for the macro-analyses. In our example (Figure 5a), both perceptual independence and decisional separability are unknown, but may hold.

If you want more specific information about the tests conducted, simply type the name of the object previously used to store the results of this analysis:

```
micro_results
```

This produces the output in Figure 5b. Results from each one of the tests summarized in Figure 5a are now displayed in full detail, including the specific subtests, computed statistics, *p*-values, and conclusions.

a.

```
> summary(micro_results)
```

| Stimulus Pair | Sampling Independence | Equal Cond d' | Equal Cond c | PI | DS |
|---|---|---|---|---|---|
| A|B1 | NO | NO | NO | no? | no? |
| A|B2 | NO | YES | NO | no? | no? |
| B|A1 | NO | YES | NO | no? | no? |
| B|A1 | NO | YES | NO | no? | no? |

b.

```
> summary(micro_results)
```

Test of Sampling Independence

| Stimulus | Response | Expected Prob | Observed Prob | Z stat | P-Value | Pass? |
|---|---|---|---|---|---|---|
| A1B1 | a1b1 | 0.49 | 0.56 | 1.57 | 0.11608 | YES |
| A1B1 | a2b1 | 0.21 | 0.14 | 2.05 | 0.04031 | NO |
| A1B1 | a1b2 | 0.21 | 0.14 | -2.24 | 0.02501 | NO |
| A1B1 | a2b2 | 0.09 | 0.16 | 2.29 | 0.02218 | NO |
| A2B1 | a1b1 | 0.27 | 0.36 | 2.14 | 0.03258 | NO |
| A2B1 | a2b1 | 0.45 | 0.36 | -2.01 | 0.04405 | NO |
| A2B1 | a1b2 | 0.23 | 0.02 | -7.80 | 0.00000 | NO |
| A2B1 | a2b2 | 0.39 | 0.26 | -3.13 | 0.00173 | NO |
| A1B2 | a1b1 | 0.25 | 0.34 | 2.17 | 0.03036 | NO |
| A1B2 | a2b1 | 0.11 | 0.02 | -4.09 | 0.00004 | NO |
| A1B2 | a1b2 | 0.21 | 0.36 | 3.77 | 0.00016 | NO |
| A1B2 | a2b2 | 0.09 | 0.28 | 5.64 | 0.00000 | NO |
| A2B2 | a1b1 | 0.04 | 0.08 | 2.15 | 0.03165 | NO |
| A2B2 | a2b1 | 0.28 | 0.24 | -1.14 | 0.25390 | YES |
| A2B2 | a1b2 | 0.10 | 0.03 | -3.07 | 0.00213 | NO |
| A2B2 | a2b2 | 0.79 | 0.65 | -3.44 | 0.00058 | NO |

Test of equal conditional d's:

| Test | d'Hits | d'Miss | Z Stat | P-value | Pass? |
|---|---|---|---|---|---|
| d_A conditional on B1 | 0.84 | 1.48 | -2.04 | 0.04162 | NO |
| d_A conditional on B2 | 1.83 | 2.26 | -1.27 | 0.20513 | YES |
| d_B conditional on A1 | 0.89 | 1.44 | -1.80 | 0.07249 | YES |
| d_B conditional on A2 | 0.83 | 1.15 | -0.70 | 0.48308 | YES |

Test of equal conditional c:

| Test | c Hits | c Miss | Z Stat | P-value | Pass? |
|---|---|---|---|---|---|
| d_A conditional on B1 | -0.01 | -1.58 | 6.03 | 0e+00 | NO |
| d_A conditional on B2 | -1.68 | -0.66 | -4.50 | 1e-05 | NO |
| d_B conditional on A1 | -0.04 | -1.50 | 6.05 | 0e+00 | NO |
| d_B conditional on A2 | -0.63 | 0.57 | -4.46 | 1e-05 | NO |

**Figure 5.** (a) Summary and (b) full results of a macro-analysis of data from a 2x2 identification design with grtools.

Recent research has found that summary statistic analyses of the 2x2 identification design can sometimes lead to the wrong conclusions about separability and independence (Mack et al., 2011; Silbert and Thomas, 2013). For this reason, it is good practice to perform the additional model-based analysis of the data that is described in the following two sections.

### 3.2 Model-based analyses using traditional GRT models

A second approach to data analysis using GRT consists on fitting a number of GRT models to the data from a confusion matrix, and then using traditional model selection approaches to determine which of those models provides the best account of the data (Ashby and Lee, 1991; Thomas, 2001). The models used implement different assumptions about perceptual separability, perceptual independence, and decisional separability. Thus, selecting a particular model is equivalent to selecting those assumptions that explain the data best.

An introductory tutorial to model fitting and selection with traditional GRT models of the 2x2 identification task can be found elsewhere (Ashby and Soto, 2015). Here, we focus on showing how to easily obtain the results from such a procedure using *grtools*.

Model-based GRT analyses have been used less often than summary statistics in the literature, probably due to the fact that implementing a variety of models, fitting them to data, and selecting among them requires quantitative skill and a deep understanding of the theory. One of the goals of developing *grtools* was to provide experimental psychologists who lack formal quantitative training with tools to easily perform and interpret model-based analyses. Thus, we wrote the software placing ease-of-use above flexibility: a full model-based analysis can be performed with only three lines of code, but the researcher is not free to choose what models to test or the procedures used for model fit and selection[1]. We start by briefly reviewing our choices regarding these points.

Because past research almost exclusively has used the 2x2 identification task (for an exception, see Ashby and Lee, 1991), *grtools* includes functions to model data only from that task. The design is popular because it has a number of advantages: it requires a relatively short experiment, it does not tax the participants' working memory as larger designs do, and it allows a study of stimulus "components" that cannot be ordered along a continuous dimension (e.g., face identity; see Soto et al., 2015).

Our focus on the 2x2 identification task means that decisional separability must be assumed throughout the analysis and cannot be tested. The reason is that it has been shown that it is impossible to perform a valid test of decisional separability using traditional modeling of the 2x2 identification task (Silbert and Thomas, 2013). However, this problem is specific to the case in which each model is fit to data from a single participant (see Soto et al., 2015), so it can be easily surpassed, as we show in the next section.

Additionally, the 2x2 identification task generates only sixteen data points per participant. Due to this small sample size, complex models (i.e., models with many parameters) cannot be adequately fit, and additional assumptions must be made to simplify all models. As in previous related research, the model-based analysis implemented in *grtools* assumes that all variances of the four perceptual distributions are equal to one.

Figure 6a shows the twelve models used in the *grtools* analysis, ordered in a hierarchy in which models that make more assumptions are placed higher, and models that make fewer assumptions are placed progressively lower. This hierarchy has been modified from one presented earlier by Ashby and Soto (2015), which in turn was based on an earlier hierarchy by Thomas (2001).

In Figure 6a, DS stands for decisional separability. As it can be seen, this property is assumed in all models. PS stands for perceptual separability and PI stands for perceptual independence. 1_RHO stands for a specific violation of perceptual independence in which the correlation (i.e., which measures the strength of the PI violation) is the same for all four perceptual distributions. At the top of the hierarchy in Figure 6a is the most constrained model, which assumes perceptual independence (PI) and separability (PS) in addition to decisional separability (DS). This model has a total of four parameters that can be varied to provide a good fit to the data (so there are four "free parameters"). Each of the models one step lower in the hierarchy relaxes a single assumption of the more general model. Relaxing an assumption requires adding one or more free parameters to the model. The process of relaxing assumptions by adding parameters continues until we arrive at the least constrained model at the bottom, which assumes only decisional separability (DS).

---

[1] However, since our software is open source and distributed under a creative commons license, any researcher can modify the original code to fit his/her own needs.

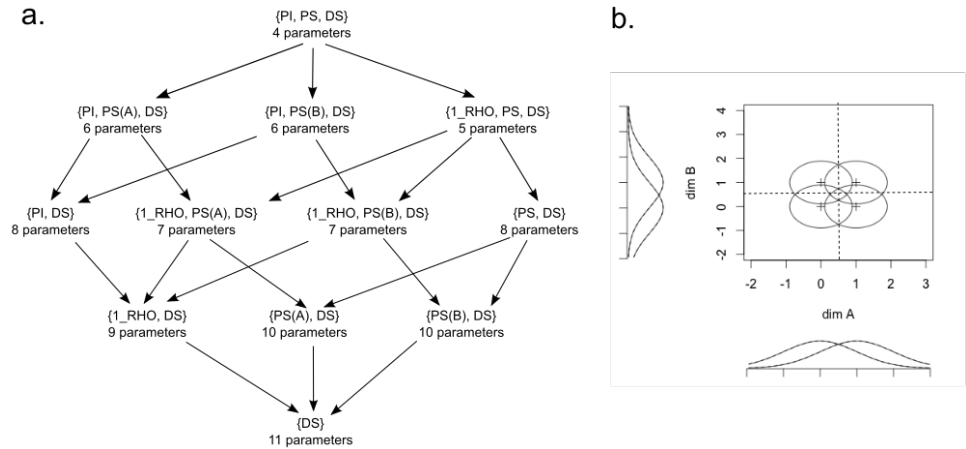

**Figure 6.** (a) Hierarchy of models used in a model-based analysis of data from a 2x2 identification task using traditional general recognition theory with grtools. PI stands for perceptual independence, PS for perceptual separability, DS for decisional separability and 1_RHO for a single correlation in all distributions. The number of free parameters in each model is indicated below its label. (b) Initial configuration assuming PS, PI and DS.

In past applications (e.g., Ashby and Lee, 1991; Soto et al., 2015; Thomas, 2001) model selection proceeded by a series of likelihood ratio tests comparing models joined by arrows in Figure 6a (for details, see Ashby and Soto, 2015). The use of the likelihood ratio tests allowed the researcher to fit only some of the models in the hierarchy to the data, speeding up the analysis. However, the increase in speed came at the cost of consistency in the model-selection procedure: in most cases, the likelihood ratio tests could not decide between several candidate models, which then had to be compared by some other criterion. Fortunately, this is no longer necessary, as the *grtools* code is fast enough[2] that the process of fitting the whole hierarchy to the data of a participant and selecting the best model can be executed in about 20 seconds. This is done through a single line of code:

```
hm_fit_results <- grt_hm_fit(cmat)
```

Here we have used again the data previously stored in the matrix `cmat`. What the function `grt_hm_fit` does in the background is fitting all the models in Figure 6 to the data in `cmat`, using maximum likelihood estimation (see Myung, 2003) through the R function `optim()`. The likelihood function of a GRT model may have several "hills" and "valleys." Ideally, the optimization algorithm would only stop at the top of the highest hill (the maximum likelihood), but it is also possible for the algorithm to stop at one of the smaller hills, which is known as a local maximum (see Ashby and Soto, 2015; Myung, 2003). To avoid this problem, the search for the best set of parameters for a particular model is performed 10 times, each time starting from a different set of initial parameter values. The best solution is kept from these 10 runs of the algorithm.

---

[2] The code in *grtools* is relatively fast for two reasons: (1) we use an efficient method of numerical integration proposed by (Ennis and Ashby, 2003) to compute the likelihood of the data for a given GRT model, and (2) we implement this method through a C++ function, which is called from R using the package Rcpp (Eddelbuettel et al., 2011).

Each set of initial parameters is determined by adding a random perturbation to a fixed configuration, shown in Figure 6b. The maximum value of the random perturbation for each parameter is set by default to 0.3.

Although the default settings for `grt_hm_fit` have worked well in our own research, users can change all of them. For example, to fit each model 20 times and use a maximum random perturbation of 0.5, one can use the following command:

```
hm_fit_results <- grt_hm_fit(cmat, rand_pert=0.5, n_reps=20)
```

Furthermore, users can have additional control over the optimization process by passing a list of control parameters as final argument to `hm_fit_results`, in the same way in which this is usually done for the `optim` function (`control=list(...)`; for more information, type `?optim` in the R console).

After finding maximum likelihood estimates for all models, the best model is selected by computing the (corrected) Akaike information criterion (AIC) of each model (Bozdogan, 2000; Wagenmakers and Farrell, 2004; for a tutorial on its application to GRT, see Ashby and Soto, 2015). A summary of the results of the model fitting and selection procedures can be obtained through the following command:

```
summary(hm_fit_results)
```

This should produce an output similar to that shown in Figure 7a. The first column of Figure 7a lists the different models tested, ordered from best to worst fit to the data. `PS` stands for perceptual separability, `PI` for perceptual independence, `DS` for decisional separability, and `1_RHO` describes a model with a single correlation parameter for all distributions. In our example, the model that fit the data best is summarized as `{1_RHO,PS(B),DS}`. This is a model that assumes a single correlation parameter (and so a violation of perceptual independence), assumes perceptual separability for dimension B, and assumes decisional separability.

The following two columns list the log-likelihood and AIC values for each of the models fitted to data. These values are used to assess which models provide the best description of the observed data. In the table produced by *grtools*, models are ranked based on the AIC value, which takes into account both the model's fit to data as well as the model's complexity. A relatively low AIC value represents a model that provides a good description of the data without being overly complex and flexible (see Bozdogan, 2000; Wagenmakers and Farrell, 2004). The final column represents the AIC weights (see Wagenmakers and Farrell, 2004), which can be interpreted as the probability that a given model is closest to the true model among those tested. In the case in Figure 7a, the probability of model `{1_RHO,PS(B),DS}` being closest to the true model is so high that it receives an AIC weight of one after rounding, with all other models receiving a weight of 0. Thus, in this case we can be very confident that model `{1_RHO,PS(B),DS}` provides the best account of the data of all the models that we tested.

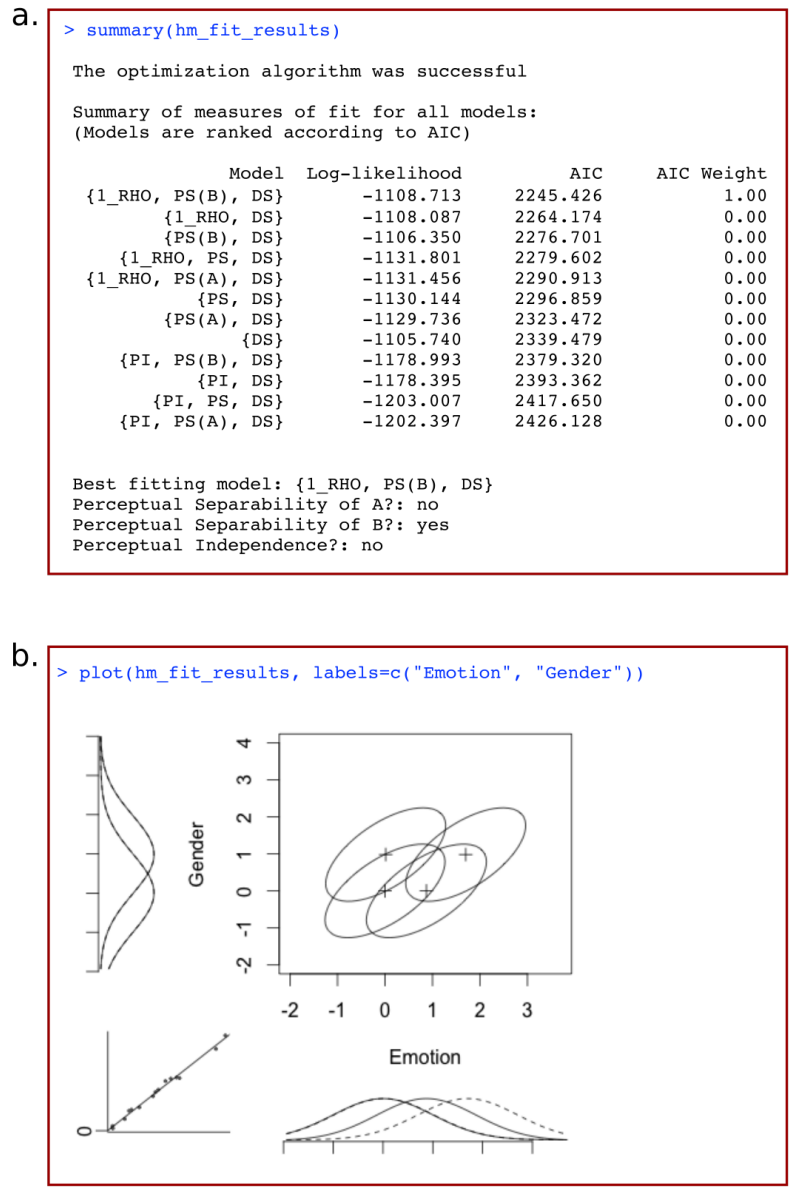

**Figure 7.** (a) Summary of the results of a model-based analysis of data from a 2x2 identification task using traditional general recognition theory with grtools, and (b) plot of the best-fitting model resulting from such analysis.

Underneath the data table, the code also provides a summary table that lists the best-fitting model, and whether perceptual separability of A, perceptual separability of B, and perceptual independence were violated.

To help visualize what the best-fit model looks like, *grtools* provides aid through the `plot` function:

```
plot(hm_fit_results, labels=c("Emotion", "Gender"))
```

This produces a graphical representation of the best-fitting model, like the one shown in Figure 7b. The elements in this figure should be interpreted as explained earlier: each ellipse represents a single Gaussian distribution, with the plus sign representing its mean. The tilt of the ellipse represents the correlation between the perceptual effects in the two dimensions. Any tilt is indicative of a failure of perceptual independence. It can be seen that the distributions show a positive correlation, indicative of a failure of perceptual independence.

Marginal distributions are plotted to the left and bottom of the figure. The marginal distributions for a given dimension are plotted using either solid or dotted lines depending on the level of the opposite dimension. These provide a visual test of whether the data suggest violations of perceptual separability: failures of perceptual separability are suggested by non-overlapping solid and dotted marginal distributions. In Figure 7b, the non-overlapping marginal distributions along the expression dimension suggest a failure of perceptual separability of expression from gender. The overlapping marginal distributions along the gender dimension suggest perceptual separability of gender from expression.

The insert at the bottom-left shows a plot of the observed response proportions against the response probabilities predicted by the best-fitting model. This plot allows a visual evaluation of how well the model fit the data. In a perfect fit, all the dots would land on the diagonal.

### 3.3 Model-based analyses using GRT-wIND

Several problems have been identified with traditional GRT model-based analyses. The most important are that traditional analyses 1) do not allow a clear dissociation between perceptual and decisional separability (Silbert and Thomas, 2013), 2) do not allow the full model to be fit to the data (because the full model has too many free parameters), and 3) are prone to over-fitting the data (Soto et al., 2015). All these problems are solved by the recently proposed General Recognition Theory with Individual Differences (GRT-wIND) model (Soto et al., 2015). GRT-wIND is fit simultaneously to the data from all participants in a particular experiment, because it assumes that some aspects of perception are common to all people. In particular, the model assumes that PS and PI either hold or do not hold across all participants, and that all participants perceive the same set of stimuli in a similar manner. On the other hand, the model assumes individual differences in decisional strategies (e.g., in whether DS holds) and in the level to which people pay more or less attention to different dimensions.

Unlike the other analyses that we have illustrated, GRT-wIND accounts for the data from all participants in an experimental group who completed the 2x2 identification task. Suppose we have data from 5 participants that we have collected in 5 confusion matrices, one for each participant. The first step in applying GRT-wIND is to take all of our confusion matrices and concatenate them in a list:

```
cmat_list <- list(cmat_1, cmat_2, cmat_3, cmat_4, cmat_5)
```

An example of properly formatted data is included with *grtools*, and can be accessed by typing the following in the R console:

```
data(grtwind_data)
```

You can look at the loaded data by typing `cmat_list` in the R console. These data have been sampled directly from the model shown in Figure 8a. This model was built to show PS of dimension B from dimension A, but failure of PS of dimension A from dimension B (marginal distributions along the *x*-axis are not aligned, see Figure 8a). There are also violations of perceptual independence (i.e., negative correlations) for stimuli $A_2B_1$ and $A_1B_2$. Furthermore, the decision strategies of hypothetical participants randomly deviated from DS on both dimensions.

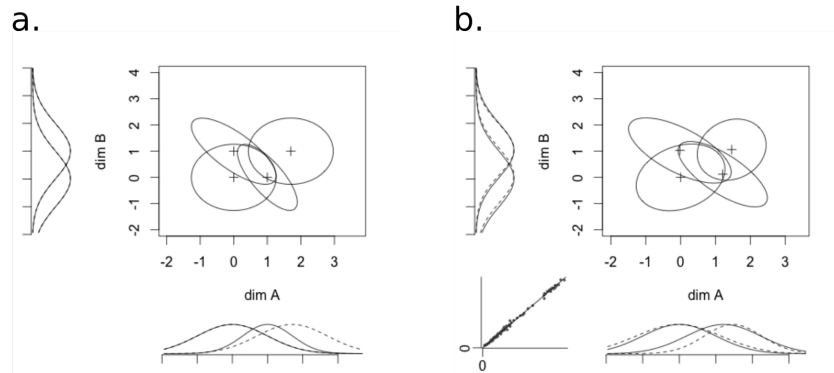

**Figure 8.** (a) Original GRT-wIND model used to generate the data in our example (see main text) and (b) model recovered by grtools.

A model-based GRT-wIND analysis starts by fitting the full model to the data in `cmat_list`, using the following command:

```
fitted_model <- grt_wind_fit(cmat_list)
```

This line of code will fit a full GRT-wIND model to the experimental data, using maximum likelihood estimation (see Soto et al., 2015). By default, the algorithm starts with a set of parameters reflecting PS, PI and DS (see Figure 6b), which are slightly changed by adding or subtracting a random value. The algorithm then efficiently searches the parameter space to determine the parameter values that maximize the likelihood of the data given the model. As indicated above, to avoid finding a local maximum, one solution is to fit the model to data several times, each time with a different set of starting parameters. This procedure is time consuming, but it provides more valid results, so we recommend using it. *grtools* includes a special function to perform such multiple fits to a GRT-wIND model:

```
fitted_model <- grt_wind_fit_parallel(cmat_list, n_reps=60)
```

`n_reps` must be provided and it represents the number of times that the model will be fit to data. In our own practice, we have chosen a rather high value of 60 for this parameter (Soto et al., 2015; Soto and Ashby, 2015). Good results (i.e., recovery of the true model producing the data) can be sometimes obtained with 20-30 repetitions. Future simulation work will be necessary to determine the minimum number of repetitions that produces good results when fitting GRT-wIND across a variety of circumstances. In the meanwhile, our recommendation is to proceed with caution and use a high value for

`n_reps`. This will considerably increase the time that the analysis will take to finish, but it will ensure more valid results.

The function `grt_wind_fit_parallel` takes advantage of multiple processing cores in the computer when these are available. By default, the function will use all the available cores minus one. This means that using a machine with multiple cores will considerably speed up the analysis.

The default settings for `grt_wind_fit` and `grt_wind_fit_parallel` have worked well for us, and we recommend that researchers with limited modeling experience use them. However, options to control the optimization algorithm and starting parameters are available for both functions. Interested researchers should read the documentation for each function, which is available by executing the commands `?grt_wind_fit` or `?grt_wind_fit_parallel` in the R console.

To visualize the best-fitting model found by the GRT-wIND analyses, you can call the `plot` function:

```
plot(fitted_model)
```

This produces Figure 8b, a representation of the recovered model that has nearly the same features as the true model shown in Figure 8a, and can be interpreted in the same way.

Additionally, you can run statistical tests to determine whether perceptual separability, decisional separability, or perceptual independence were violated. Currently, *grtools* offers two ways to perform such statistical tests: Wald tests (see Soto et al., 2015) and likelihood ratio tests (see Ashby and Soto, 2015).

The Wald tests have the advantage of being relatively fast to compute. However, they require an estimate of the Hessian matrix of the likelihood function associated with the maximum likelihood estimates (see appendix of Soto et al., 2015), which is computationally costly and therefore may take several minutes. Even so, once the Hessian is obtained, all tests can be computed in seconds.

Unfortunately, some estimates of the Hessian are not useful for the computation of a Wald test (i.e., when the Hessian is not positive definite, see Gill and King, 2004). In our experience, the estimates provided by R (e.g., through the package "numDeriv", see Gilbert and Varadhan, 2015) are often problematic. We are currently working on implementing procedures to estimate the Hessian that have proven more successful in our experience (e.g., DERIVEST; D'Errico, 2006).

Because of these problems with the Wald test, we recommend researchers perform likelihood ratio tests instead. These tests are slow to compute, but they do not require numerical estimation of the Hessian. A full series of likelihood ratio tests for PS, PI and DS is performed by using the following line of code:

```
fitted_model <- lr_test(fitted_model, cmat_list)
```

Each likelihood ratio test requires refitting GRT-wIND to the data an additional time, except this time with a model in which the assumption being tested (e.g., PS of dimension A) is assumed to hold. As before, this process is repeated a number of times with different starting parameters, obtained by adding random perturbations to the parameters estimated earlier with `grt_wind_fit_parallel`. Because these parameters are likely to be a better starting guess than the model in Figure 6b, for most cases we can obtain good results without having to run a high number of repetitions. By default, the function `lr_test` runs the optimization algorithm 20 times for each test. This value can be changed by explicitly setting `n_reps` to a different value, as you did in the call to `grt_wind_fit_parallel`.

Using **summary**`(fitted_model)` should now print to screen an output similar to that shown in Figure 9. There are three parts to this output. The first line is a message indicating whether the optimization algorithm was successful in finding the maximum likelihood estimate of the GRT-wIND parameters. Any problem in the optimization process (which may invalidate the results of this analysis) will be described in this line. This is followed by a summary of the fit of the full model to data, including both the obtained log-likelihood and R-squared (the proportion of the variance in the data explained by the model) of the best-fitting model. The third part of this summary output includes the results for all likelihood ratio tests, in a table with columns including a description of the test, the Chi-squared test statistic, degrees of freedom, p-value, and the test's conclusion (i.e., whether or not PS, PI or DS is violated). In our example, the likelihood ratio tests accurately conclude that PS holds for dimension B (i.e., NO violation) but fails for dimension A (YES violation), that PI does not hold (YES violation), and that DS does not hold for either dimension (YES violation).

```
> summary(fitted_model_test)

The optimization algorithm was successful

Measures of Fit:
 Log-likelihood: -6524.54
 R-squared: 0.9967

Results of the Likelihood Ratio tests:

                                  Test   Chi2 DF   pval  Violation
Perceptual Separability of dimension A  18.75  4  0.001        YES
Perceptual Separability of dimension B   1.07  4  0.899         NO
                Perceptual Independence  26.95  4  0.000        YES
Decisional Separability of dimension A  16.65  5  0.005        YES
Decisional Separability of dimension B  12.03  5  0.034        YES
```

**Figure 9.** Summary of the results of a model-based analysis of data from a 2x2 identification task using general recognition theory with individual differences (GRT-wIND).

If only some likelihood ratio tests are of interest, it is possible to explicitly indicate what tests to run:

```
fitted_model <- lr_test(fitted_model, cmat_list test=c(
"PS(A)", "PS(B)", "PI", "DS(A)", "DS(B)" ))
```

To run only some of these tests, simply keep the strings in the `test` array that correspond to a test that you want to run, and delete other strings. For example, to run tests of perceptual and decisional separability of dimension A only, you should use `test=c("PS(A)", "DS(A)")`.

## 4. The 2x2 Garner Filtering Task

Along with the 2x2 identification task, *grtools* also provides tools to analyze data gathered from a related task, known as the 2x2 Garner filtering task (Garner, 1974). The stimuli are identical to those used in the identification task, but the Garner filtering task asks participants to classify stimuli based on their value on a single dimension, while ignoring stimulus values on the second dimension. For example, a 2x2 Garner filtering task may require participants to classify faces according to their gender while ignoring their emotional expression (gender task), or to classify faces according to emotional expression while ignoring their gender (emotion task). This differs slightly yet significantly from the identification task; instead of identifying a unique stimulus, participants are now asked to group the stimuli based on a single dimension (e.g., gender or emotion).

There are two important conditions in a Garner filtering task, presented to participants in separated experimental blocks. In baseline blocks, the value of the irrelevant dimension (the dimension that participants must ignore) is fixed across trials. For example, participants may have to classify faces according to their gender, while emotional expression is fixed at "happy." In filtering blocks, the value of the irrelevant dimension is varied across trials. For example, participants may have to classify faces according to their gender, while emotional expression randomly varies between "happy" and "sad."

If participants have difficulty separating one dimension from another, then a "Garner interference effect" (Garner, 1974) is expected: slower response times and lower accuracy are likely to occur on filtering blocks, compared to baseline blocks. Accuracy in the Garner filtering task is usually at ceiling levels, so most studies have measured the Garner interference effect in terms of response times.

### 4.1 Summary statistics analysis

The Garner filtering task is widely known and used in perceptual and cognitive psychology, with most applications focusing on the Garner interference effect as a test for dimensional separability. Because response times are the most common dependent variable used for the filtering task, using GRT for data analysis requires making further assumptions about the way in which parameters in a GRT model are related to observed response time distributions. By making the simplest possible assumption (the *RT-distance hypothesis*; see Murdock, 1985), Ashby and Maddox (1994) showed that the data from a filtering task can be used to compute several tests of dimensional separability that are *more diagnostic* than the traditional Garner interference test.

In particular, whereas a violation of separability (perceptual or decisional) is likely to produce a Garner interference effect, a *context effect* is also likely to produce an interference effect (Ashby and Maddox, 1994). Context effects refer to the case in which the perception of a particular stimulus is changed depending on other stimuli presented closely in time. In the Garner task, such a change in context happens between the baseline (two stimuli are presented) and the filtering (four stimuli are presented) blocks. The effect of context on stimulus perception does not need to be related to separability at all. For example, variability in the irrelevant dimension during filtering blocks could produce spontaneous switches of attention towards that dimension, which are quickly remedied by switching attention back to the relevant dimension (see Yankouskaya et al., 2014). Such switches of attention would increase response times in the filtering blocks compared to the baseline blocks, even when dimensions are perceptually separable.

Ashby and Maddox (1994) proposed two tests of separability that are not affected by context effects: marginal response invariance (mRi) and marginal response time invariance (mRTi) tests. Context effects do not influence mRi and mRTi because these tests are computed from data originating from a single block type. Usually, the data should come from filtering blocks, because in these blocks all stimuli are presented at the same time, which reduces the likelihood of participants changing their response strategy depending on the specific stimuli presented.

The mRi test consists on comparing the proportion of correct responses for the relevant dimension across the two levels of the irrelevant dimension. If both PS and DS hold, then the mRi test should indicate no significant differences across levels of the irrelevant dimension. For example, mRi for gender across changes in emotion means that the probability of correctly classifying a face as male does not change depending on whether the face is sad or happy.

The mRTi test compares the full distribution of response times for correct classifications of the relevant dimension across the two levels of the irrelevant dimension. If both PS and DS hold, then the mRTi test should indicate no significant differences across levels of the irrelevant dimension. For example, mRTi for gender across changes in emotion means that the time it takes to correctly decide that a face is male does not change depending on whether the face is sad or happy. Because mRTi involves a comparison of whole response time distributions, it is the strongest test of separability available for the Garner filtering task (Ashby and Maddox, 1994).

*grtools* includes a routine to perform all the summary statistic analyses for the 2x2 Garner filtering task described above using only a couple of commands. Data are analyzed for each participant individually, and the first step in the analysis is to create a data frame with the participant's data. Each row in the data frame includes data from one trial, with the following column organization:

- Column 1: block type, with a value of 1 for baseline blocks and a value of 2 for filtering blocks
- Column 2: Level of the relevant dimension, with values 1 and 2
- Column 3: Level of the irrelevant dimension, with values 1 and 2
- Column 4: Accuracy, with a value of 0 for incorrect trials and 1 for correct trials
- Column 5: Response time

The data can be entered into a data spreadsheet processor like Microsoft Excel, saved as a comma-separated file, and then imported to R using the function `read.table`, as explained for `cmat` above. For this analysis however, the imported data should not be converted to a matrix. An example of properly formatted data is included with grtools, and can be accessed by typing the following in the R console:

```
data(garner_data)
head(garner_data)
```

The first line of code loads the `garner_data` data frame into our current R session, and the second line of code prints the column names and the first few rows on the console. We perform the summary statistic analysis on these data using the following line of code:

```
garner_results <- sumstats_garner(garner_data)
```

a.
```
> summary(garner_results)

                    Separability Test   Pass?
   Garner Interference - Accuracy       yes?
        Garner Interference - RT        yes?
    Marginal Response Invariance        yes?
          Marginal RT Invariance        yes?
```

b.
```
> garner_results

Garner interference test:

               Test  Baseline  Filtering  Difference  Statistic  P_value  GI effect?
  Proportion Correct     0.88       0.83      -0.052        1.51  0.13112          NO
   Median correct RT     0.77       0.70      -0.073    17986.50  0.99592          NO

   Note: Uses Wilcoxon test to compare RTs and a z-test to compare
   proportions in the baseline and filtering conditions

Marginal response invariance test:

                             Test   Z stat   P-value   Pass?
   Level 1 of relevant dimension    -1.34    0.17962    YES
   Level 2 of relevant dimension     1.55    0.12071    YES

Marginal response time invariance test:

          RT distributions from correct trials
                     Test  KS stat   P-value   Pass?
  Level 1 of Relevant Dimension  0.1980   0.33711    YES
  Level 2 of Relevant Dimension  0.1943   0.37830    YES

   Note: Uses Kolmogorov-Smirnov test to compare RT distributions
   across the two levels of the irrelevant dimension
```

**Figure 10.** (a) Summary and (b) full results of a summary-statistics analysis of data from a 2x2 Garner filtering task with grtools.

The following line of code provides a comprehensive summary of the results:

**summary**(garner _results)

This produces the output shown in Figure 10a. The first column lists the name of the separability test, while the second column shows whether the test was passed (yes?) or not (NO) according to the analysis. The statistical test used by default to compare response time distributions in the mRTi test is the well-known Kolmogorov-Smirnov test, or KS test. If the conclusions from these analyses are that all tests are passed (yes?), then this suggests that the relevant dimension might be separable from the irrelevant dimension. If any of the conclusions from these analyses indicate that the test is not passed (NO), then this suggests that the relevant dimension is not separable from the irrelevant dimension. As with traditional GRT analyses of the identification task, but unlike analyses based on GRT-wIND, the summary-statistics analyses of the Garner interference task cannot determine whether violations of separability are due to perceptual or decisional factors. The usual approach is to assume that DS holds and make conclusions about PS. As indicated in section 4 below, there are ways to design the task so that DS is more likely to hold.

To view a more detailed description of the results of this analysis, simply call the R object in which you stored the results:

`garner_results`

This produces the output shown in Figure 10b. Full details and information of the analysis are shown, including the specific subtests, computed statistics, *p*-values, and conclusions.

Although model-based analyses of data from the Garner filtering task are possible (Maddox and Ashby, 1996), they require previous knowledge about the perceptual distributions involved, which can be acquired by fitting a GRT model to the data from an identification task. That makes the analyses rather redundant, so applications of GRT to the filtering task have not used model-based analyses.

## 5. Some general recommendations

We would like to finish by providing some general recommendations regarding design and analyses aimed at determining dimensional independence and separability:

1. *For best results, run a 2x2 identification experiment and analyze it using GRT-wIND.* The 2x2 identification task is appealing due to its simplicity and because it is easy to run. Currently, model-based analysis with GRT-wIND is the only way to dissociate between perceptual and decisional processes in the 2x2 identification task. Although the Garner filtering task is very popular among researchers, this task does not allow one to dissociate perceptual and decisional types of separability, as does the identification task.

2. *Calibrate stimulus values in the identification task to ensure that all participants produce a moderate rate of errors.* Errors are the only source of information for the GRT analyses of the identification task presented here. If a participant in your experiment does not make errors, then that participant does not contribute any information about dimensional separability and independence. The opposite is also true: if a participant shows performance near chance levels, then that participant also does not contribute information to the analysis. For this reason, you should calibrate your stimuli to make sure that participants show a moderate error rate (~25%-35%). Common ways in which this can be achieved are by changing presentation times, contrast levels, and morphing levels (for faces and other objects). Usually, running some pilot participants is enough to settle on a particular set of stimuli. One possibility– that we have not explored yet–is using adaptive procedures (for reviews, see Leek, 2001; Lu and Dosher, 2013) to bring the error rates of all participants to the same level, by varying some stimulus feature that is unrelated to the dimensions under study.

3. *If you have the resources, use a concurrent operations approach by also running a 2x2 Garner filtering task.* In theory, the results from this second task cannot give more information about dimensional independence than the results from the identification task. In practice, it is advantageous to verify that the results obtained with the identification task, and analyses based mostly on choice proportions, are in line with results obtained with the Garner filtering task and analyses based mostly on response times.

4. *Use instructions that will increase the likelihood of decisional separability.* Many of the analyses described here are only valid if decisional separability holds. Fortunately, experimental results have shown that there are ways to encourage participants to use such decisional strategies, such as providing instructions indicating that stimuli follow a grid configuration (e.g., a 2×2 grid) and spatially positioning response buttons in that configuration (Ashby et al., 2001).

5. *Use all available tests of separability for the Garner filtering task.* The most common test of separability for the Garner task is checking whether interference occurs by comparing performance in baseline and filtering blocks. As we have seen before, two other tests of separability are available for the Garner task – mRi and mRTi– and both are more diagnostic about dimensional separability than the Garner interference test. Given that there is no additional cost to perform such tests, we believe that most researchers do not perform them because they simply do not know how. Fortunately, *grtools* allows all three tests to be performed using a single command.

6. *Include a control group that provides a benchmark of separability*. Violations of separability are common, and features of your task and design might produce such violations. This is particularly troublesome when using a design and analysis that does not distinguish between perceptual and decisional forms of separability, such as the Garner filtering task or traditional analyses of the identification task. However, we now know that factors such as training in a categorization task can influence perceptual separability as well (Soto and Ashby, 2015). For these reasons, it is a good idea to include a group that would serve as a benchmark, if the design does not already involve more than one group. That benchmark could be very simple; examples include a group presented with dimensions known to be separable (e.g., orientation and width of gratings, shape and color) or integral (e.g., brightness and saturation; unfamiliar face identities).

7. *Make sure that your results are generalizable, rather than explainable by task features*. The simplicity of the 2x2 tasks comes at a cost: only four stimuli are studied, although the number of combinations of levels of the two dimensions under study could be infinite. You should be careful to not over generalize when interpreting the results from a single set of stimuli. Ideally, several experiments should be performed before reaching a conclusion, perhaps parametrically varying task factors such as difficulty (Yankouskaya et al., 2014). A good and low-cost way of learning whether task features can explain performance in a visual task is by analyzing the performance of an ideal observer in the task (for a GRT analysis of an ideal observer's performance, see Experiment 2 of Soto and Wasserman, 2011).

## 6. Conclusion

We have described an R package with functions that perform a variety of statistical analyses to determine independence and separability of perceptual dimensions according to GRT. In the past, GRT has been popular among mathematical psychologists who have the technical knowledge to implement statistical analyses using it. Most of the research using GRT outside this community was facilitated by Kadlec's publication of the *mdsda* program (Kadlec, 1995, 1999), which performed summary statistics analyses of data from the identification tasks. Recently, another R package has been made public that allows data analyses using traditional GRT models (Silbert and Hawkins, 2016). However, this package requires a more active involvement from the researcher during the analysis. The researcher must have a good understanding of the relation between a variety of tests and the concepts of PI, PS and DS. For model-based analyses, the researcher must decide what models to fit and compare, and make a selection based on measures of fit. On the other hand, *grtools* was developed with the typical experimental psychologist in mind, someone who wants to make use of the sophisticated analytical tools offered by GRT, but does not have the training to implement the analyses, make critical decisions about procedures, and interpret the overall pattern of results. Our package allows researchers to perform full GRT analyses with only three commands, and explore the results in a way that highlights the most important conclusions from the study. Importantly, *grtools* is the only currently available package implementing summary statistics analyses of the widely-used Garner filtering task and model-based analyses with GRT-wIND (Soto et al., 2015). The latter are the only analyses capable of dissociating perceptual and decisional forms of separability with a 2x2 identification design. We hope that the availability of *grtools* will lead to a wider application of GRT to the analysis of dimensional interactions.